\documentclass{article}
\usepackage{authblk}
\author[1]{Camila Díaz}
\author[2,1]{Jocelyn Dunstan}
\author[3]{Lorena Etcheverry}
\author[2]{Antonia Fonck}
\author[1]{Alejandro Grez}
\author[2,1]{Domingo Mery}
\author[2,1]{Juan Reutter}
\author[4,1,5]{Hugo Rojas}

\affil[1]{IMFD, Chile}
\affil[2]{Pontificia Universidad Católica de Chile, Chile}
\affil[3]{Universidad de la República, Uruguay}
\affil[4]{Universidad Alberto Hurtado, Chile}
\affil[5]{VioDemos, Chile}

\usepackage{amsmath}
\usepackage{amsfonts}
\usepackage{amssymb}
\usepackage{listings}
\usepackage{geometry}
\usepackage{url,multirow}

\usepackage[T1]{fontenc}

\usepackage{tcolorbox}

\tcbset{
    mypromptstyle/.style={
        colback=gray!10!white,
        colframe=black,
        fonttitle=\bfseries,
        coltitle=black,
        boxrule=0.5mm,
        arc=2mm,
        width=\textwidth,
        before=\par\vspace{10pt},
        after=\par\vspace{10pt},
        title=Prompt,
    }
}

\usepackage{xcolor}

\newcommand{\af}[1]{{\color{black}{#1}}}
\newcommand{\jd}[1]{{\color{black}{#1}}}
\newcommand{\lore}[1]{{\color{black}{#1}}}
\newcommand{\dm}[1]{{\color{black}{#1}}}

\begin{document}



\title{Automatic knowledge-graph creation from historical documents: The Chilean dictatorship as a case study}






\maketitle

\begin{abstract}
We present our results regarding the automatic construction of a knowledge graph from historical documents related to the Chilean dictatorship period (1973-1990). 
Our approach consists on using LLMs to automatically recognize entities and relations between these entities, and also to perform resolution between these sets of values. In order to prevent hallucination, the interaction with the LLM is grounded in a simple ontology with 4 types of entities and 7 types of relations. 

To evaluate our architecture, we use a gold standard graph constructed using a small subset of the documents, and compare this to the graph obtained from our approach when processing the same set of documents. Results show that the automatic construction manages to recognize a good portion of all the entities in the gold standard, and that those not recognized are mostly explained by the level of granularity in which the information is structured in the graph, and not because the automatic approach misses an important entity in the graph. 
Looking forward, we expect this report will encourage work on other similar projects focused on enhancing research in humanities and social science, but we remark that better evaluation metrics are needed in order to accurately fine-tune these types of architectures.  
\end{abstract}

\section{Introduction}

Knowledge graphs have been identified as a promising tool for analyzing historical documents \cite{gutierrez2021knowledge,debruyne2022creating,opitz2018induction}. Indeed, given a collection of documents, one can build a knowledge graph by identifying all relevant entities and relations between them. The construction and usage of such knowledge graphs allow shifting focus from a document-centric approach, in which analysts must find their information within a collection of documents, to an entity-centric approach, in which users can immediately find relevant entities in knowledge graphs, together with other helpful information such as the neighborhood of these entities, the relation or paths between them, and other more complex patterns. 

While promising, the construction of these knowledge graphs is a challenging, expensive endeavor. To identify entities, one must read all relevant documents, and this list must be constantly curated to avoid duplication. The same applies to the discovery of the relations between entities. One way of partially preventing the cost of constructing knowledge graphs is to look for automatic or semi-automatic techniques in which one leverages recent advances in Natural Language Processing \jd{(NLP)} to take care of the construction of the knowledge graph or to deliver a preliminary result that is then later manually curated at a much lower cost.  

\jd{In this paper, we present results regarding the automatic construction of a knowledge graph containing information about the human rights violations committed by the Chilean dictatorship of Augusto Pinochet between 1973 and 1990. The transition to democracy and transitional justice began in Chile after the peaceful return to democracy in 1990. In these 34 years, there has been notable progress in the five different elements of transitional justice—the search for truth, justice, reparation, memory, and non-repetition—although important challenges remain pending\cite{rojas2022human, corral202350}.} 

\af{For the work presented here, we use historical documents retrieved from the digital archive \url{memoriaviva.org}. This repository is part of the NGO Human Rights International Project, initially established by Chilean refugees and human rights activists in London. Their goal is to collect and make available to a broad public the documentation that records the crimes committed by the dictatorial state apparatus.}


\jd{Our ultimate goal is the creation of innovative computational methods specifically designed for analyzing historical documents, which will enable the development of tools to integrate currently fragmented information while adhering to necessary quality and standards. This endeavor is also intended to impact the process of building historical knowledge, providing a space for the interrelation between fragments that enable more integrated and comprehensive analyses, supporting the work of different disciplines and organizations.}



\section{Related work}
\label{sec:related}

Information extraction tasks are usually divided into open and closed tasks. Open information extraction (OIE) is designed to derive relation triplets from unstructured text by directly using entities and relations from the sentences without adhering to a fixed schema. In contrast, closed information extraction (CIE) focuses on extracting factual data from text that conforms to a predetermined set of relations or entities, as detailed in \cite{Josifoski2021}. Our approach can be classified under the CIE paradigm, but we use a simple domain ontology to guide the process of extracting entities and relationships between entities instead of a predefined set of concrete entities or relations.

Regarding the machine learning models to extract entities and relationships, several papers in recent years have explored comprehensive methods that use a single machine learning model for joint-named entity recognition and relationship extraction (NERRE) \cite{xu2023large}. These methods typically take a sequence-by-sequence approach in which a model is trained to obtain tuples of two or more named entities and the relationship label belonging to the predefined set of possible relationships between them. In \cite{Dagdelen2024}, the authors present an approach using LLMs for NERRE in the materials science domain, extracting hierarchical entities and relationships between them. Our work with historical documents presents new challenges to this task: the documents we process contain various information,  ranging from judicial processes to family relationships. This means the extraction process must be general enough to accommodate general information instead of focusing on a particular domain. 

Although open and closed information extraction vary, both attempt to convert unstructured text into structured knowledge, usually represented in triples. These triples help outline relationships but offer limited entity-level knowledge. It is often assumed that two triples refer to the same entity if their topics coincide. However, this assumption only sometimes holds. Furthermore, evaluating these tasks is based on precision, recall, and F1 at the triplet level, which can lead to erroneous conclusions about entity understanding. 
Some recent works propose new metrics to evaluate the quality of the results obtained beyond these classical metrics. Among them, the one defining the Approximate Entity Set OverlaP (AESOP) metric stands out \cite{Wu2024}. 
In this paper, we take inspiration from this work to find metrics suitable for comparing graphs produced by two possibly different black-box architectures. We only allow comparison at the level of the final answer.

\section{Our approach}

\begin{figure}[h]
\centering 
\includegraphics[width=1.0\textwidth]{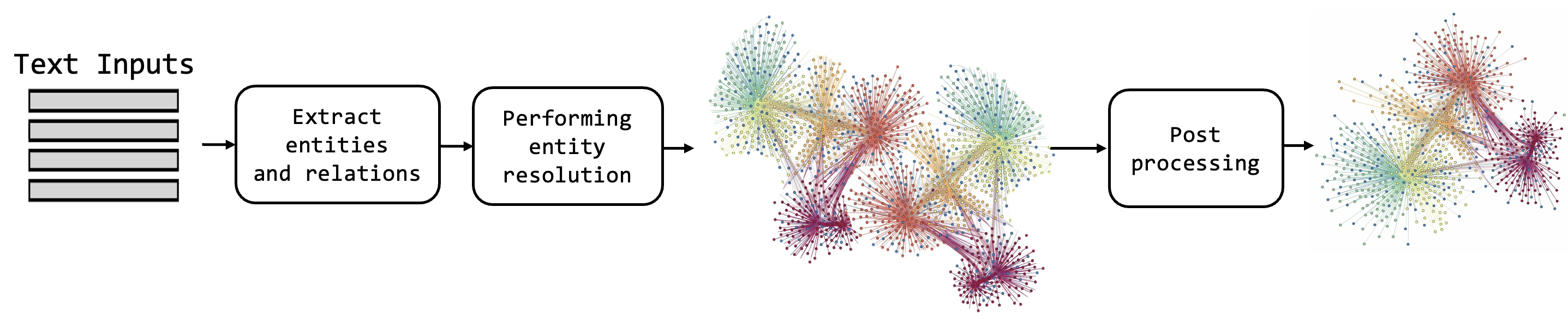}
\caption{\dm{Summary of the proposed approach.}}
\label{Fig:GeneralSchema}
\end{figure}

Our approach is summarized in Fig. \ref{Fig:GeneralSchema}. The idea is to repeatedly prompt an LLM for entities and relations and resolve these relations. This results in a graph, which is later post-processed to remove additional redundancy and fix possible mistakes in the previous steps. 

\subsection{A simple (fixed) ontology}

Our architecture considers a simple ontology in which we fix certain types of entities and relations. We do this to avoid excessive hallucination from the LLM. However, we keep the ontology general enough to avoid losing valuable information. \dm{The proposed entities and relations are presented in Table \ref{Tab:Ontology}}. 
Furthermore, each relation type also restricts the type of its origin and target entities. In the rest of the section, we omit clear types from the context.

\begin{table}[t]

\begin{minipage}[t]{0.25\linewidth}
\centering
\begin{tabular}{| l |}
\hline
\textbf{Entities} \\
\hline
\emph{Event} \\
\emph{Individual} \\
\emph{Location} \\
\emph{Organization} \\
\hline
\end{tabular}
\end{minipage}
\hfill
\begin{minipage}[t]{0.65\linewidth}
\centering
\begin{tabular}{| l | l l |}
\hline
\textbf{Relations} & \textbf{From} & \textbf{To}\\
\hline
\emph{IndividualIsRelatedToOrganization} & \emph{Individual} & \emph{Organization}\\
\emph{IndividualIsRelatedToEvent} & \emph{Individual} & \emph{Event} \\
\emph{OrganizationWasPresentAtLocation} & \emph{Organization} & \emph{Location}\\
\emph{OrganizationIsPartOfOrganization} & \emph{Organization} & \emph{Organization} \\
\emph{OrganizationIsRelatedToEvent} & \emph{Organization} & \emph{Event}\\
\emph{EventOccursAtLocation} & \emph{Event} & \emph{Location} \\
\emph{LocationIsContainedInLocation} & \emph{Location} & \emph{Location} \\
\hline
\end{tabular}
\end{minipage}
\label{Tab:Ontology}
\caption{Proposed Ontology}
\end{table}

\subsection{Extracting Entities and Relations}

Entities and relations are extracted by prompting LLMs. We use OpenAI's API\footnote{We use the GPT-4o-mini model, but if the model takes too long or repeatedly fails for one prompt, we use GPT-4o for that particular prompt and document \dm{(see details on \url{https://platform.openai.com/docs/models})}}.
\lore{The approach used is zero-shot prompting, as no examples of the expected output are included. However, the expected behavior and expected structure are detailed in a JSON document.}
 We extract each type of entity or relation independently, with a different prompt that broadly specifies the context of the documents, the data expected from the response, and the expected structure. 

For example, the prompt to recognize entities of type \textit{Individual} is as follows:

\begin{tcolorbox}[mypromptstyle, title= Prompt to extract Individuals]
\textbf{Prompt:} {\tt Your goal is to identify all the individuals mentioned in the document and provide the information about that person as a structured object. You will receive a document related to the Chilean dictatorship of 1973.

Generate a new JSON object containing the identified individuals:}

\begin{lstlisting}[basicstyle=\ttfamily\small, breaklines=true]
{
    "individual": [ // a list of all the individuals
        {    
            "firstName": string, // first name of the individual
            "lastName": string, // last name of the individual
            "role": string, // individual's job, role, profession or activity. If not specified, make it "unspecified"
            "summary": string, // 1-sentence summary of the individual
            "origin_reference": string[] // If the document has parts identified with the field "ORIGIN_REFERENCE", here comes that value; if more than one corresponds, add them all as a list of strings
        }
    ]
}
\end{lstlisting}
{\tt Make sure that each individual found satisfies this sentence: "individual person with first and last name". Before adding the individual to the result, imagine a detailed explanation for why you deduce that the individual satisfies the sentence. If the explanation is not 100\% convincing, ignore that result. Use only lowercase letters without accents. Use the English language for the summary.}
\end{tcolorbox}

As per relations, recall that we restrict each relation to the types of entities where they participate. Then, for every relation type $R$ between entities of type $E_1$ and $E_2$, we form a prompt that concatenates all entities of type $E_1$ and $E_2$ discovered in the previous process and asks the LLM to discover relations between these entities only. 
For example, here is the prompt for the relation of type \textit{IndividualIsRelatedToOrganization} (to which we must add the list of entities of type Individual and Organization that were previously discovered in the document):

\begin{tcolorbox}[mypromptstyle, title= Prompt to extract Individuals]
    \textbf{Prompt:} {\tt You will be given a JSON with a list of individuals (denoted between tags '==LIST 1 START==' and '==LIST 1 END=='), a JSON with a list of organizations (denoted between tags '==LIST 2 START==' and '==LIST 2 END=='), and then a document (denoted between tags '==DOCUMENT START==' and '==DOCUMENT END=='). Based on the context, you can identify from the document, find the individuals that are related to an organization somehow. The result must be a JSON object with the following structure:}
    
    \begin{lstlisting}[basicstyle=\ttfamily\small, breaklines=true]
    {
        "IndividualIsRelatedToOrganization": [
            {
                "nature": string, // The nature of the relation of the individual with the organization, it must be "affected by", "member", "chief", or "other"
                "individualId": number, // ID of the individual that is a member of the organization
                "organizationId": number, // ID of the organization
                "summary": string, // 1-sentence summary of the relation
                "origin_reference": string[] // If the document has parts identified with the field "ORIGIN_REFERENCE", here comes that value; if more than one corresponds, add them all as a list of strings
            }
        ]
    }
    \end{lstlisting}
    {\tt Each relationship should include a summary describing the nature of the relationship between the entities. Ensure meaningful relations are included, avoiding duplicates. Use only lowercase letters without accents. Use the English language for the summary.}
\end{tcolorbox}

\subsection{The full text-to-graph pipeline}

In the following, we explain the entire pipeline that we use to go from text to a knowledge graph, which we divide into three main steps. First, we need to split the text documents into blocks of smaller length to fit into the prompt we send to OpenAI's API. Then, we create a prompt for each fragment, entity type, and relation to extract the data from the fragment. Lastly, we process this entity/relation graph to remove redundant or incorrect information based on the graph's structure.

\subsubsection{Document Splitting}

Since we use OpenAI's GPT API, which has a size limit for its prompts input and output, we need to divide the text document into fragments of a smaller size. Moreover, we must choose a size that produces a good enough response. Generally, with a size that is too small, the AI may not know what to look for and may produce forced or incorrect results. Conversely, with a size that is too big, the response may include only some occurrences. We tested lengths of 1000, 2000, 5000, 10000, and 15000 characters, making each cut either at a line break or a sentence end. We also added an overlay at the beginning and end of each fragment of 0.1 of its size to provide the AI with more context. After trying different values, we found that the best results were obtained with fragments of 5000 characters with overlays of 500 characters.

\subsubsection{Prompting to extract Entities and Relations}

After splitting the documents into fragments, the next step is to extract the entities and relations of each fragment using OpenAI's GPT API. As we explain, to extract the entities of a fragment for each entity type, we prompt the API with the corresponding prompt as the system role and the fragment as the user role and receive a JSON object with the entities of that type found in that fragment. Similarly, to extract the relations of a fragment, we prompt the API with the corresponding prompt for each relation type and add all entities found in that fragment that are of the types relevant to the relation, i.e., the types of the origin and target entities. This is why we consider only one origin type and one target type for each relation. With this prompt, we get a JSON object with the relations of that type found in that fragment.

\subsubsection{Performing Entity Resolution}
Next comes the task of removing duplicates and filtering out incomplete entities. Once again, we resort to LLMs, feeding them the list of entities and a different prompt for each type;  each type is processed with specific criteria. For example, suppose two individuals have the same complete name or a full name containing the other. In that case, they are considered the same individual, and the more informative name is maintained.

We do the same for relations, with two additional rules: the first two relations of the same type connect the same entities and are considered duplicates. Second, specifically for relations between entities of type \emph{Location}, we remove any containment relation between locations that do not follow the order country > city > street > building. For this, we ask in the prompt of the location that any location must come with a nature property that is either one of these values or undefined.

After this, we obtain our raw graph.

\subsubsection{Graph Post-Processing}

The raw graph is also post-processed to remove incorrect and redundant information by applying rules to its structure. 

First, we focus on edge removal. Generally, it would not make sense to have loops in this kind of graph, so we remove any existing loops (we use a loop limit of size 5 to ease the computation). The next step is to remove redundant edges that can be easily extracted from our ontology. For this, we define four rules:

\begin{itemize}
    \item If \( O1 \text{ IsPartOf } O2 \) and \( O2 \text{ IsPartOf } O3 \), then \( O1 \text{ IsPartOf } O3 \) is redundant;
    \item If \( E \text{ OccursAt } L1 \) and \( L1 \text{ IsContainedIn } L2 \), then \( E \text{ OccursAt } L2 \) is redundant;
    \item If \( I \text{ WasPresentAt } L1 \) and \( L1 \text{ IsContainedIn } L2 \), then \( I \text{ WasPresentAt } L2 \) on the same date is redundant;
    \item If \( O1 \text{ IsRelatedTo } E \) and \( O1 \text{ IsPartOf } O2 \), then \( O2 \text{ IsRelatedTo } E \) is redundant.
\end{itemize}

All redundant relations can safely be removed. Note that these redundancies come from the fact that the relation types \textit{IsPartOf} and \textit{IsContainedIn} are transitive.

After removing incorrect and redundant edges, we focus on merging redundant nodes \footnote{The AI annotates every entity with a summary property, and every time a pair of entities are merged, the resulting one maintains both summaries in a list. The same is done for relations.}

Our initial experiments showed that the LLM sometimes confuses organizations and locations, resulting in duplicate entities that must be resolved.
Therefore, we address this issue by identifying locations and organizations with the same name and resolving them into one entity: a location, an organization, or both.
The latter case makes sense in specific instances, such as when an organization is located in a particular building, and the name could refer to both the organization and the building.
We decide this based on which node has more information.
The API annotates organization and location entities with a possibly unspecified \textit{nature} parameter.
If one node has this parameter defined and the other does not, then that node's type is prioritized; if both nodes have the value or neither does, then the node keeps both types.

The last step of our process is to merge duplicate event nodes.
We aim to reduce the number of event nodes by applying a limit to its granularity in two ways.
Note that every event is annotated with a property \textit{date} that contains the date it takes place.
First, we restrict that there can only be one event per location per date, so all events with the same values are merged into one.
Second, an individual can be related to only one specific event on a particular date, so all events that share these values must be merged.
Since applying any of the two rules might raise new ones for the other, we use them iteratively until no changes are made to the graph.

\section{Evaluation}


To validate the proposed approach 
 we propose to examine documents from public records collected from a known collection (in our case, \url{memoriaviva.org}). We asked a domain expert to construct a gold-standard graph from a subset of the documents in this collection, and we compared how we fare against this standard graph.


\subsection{A gold standard sub-graph}

Our gold standard graph relates to the facts presented in \url{memoriaviva.org} concerning the location of Lonquen, Chile, wherein a series of corpses were found during the dictatorship. The constructed graph contains 121 entities: 51 Individuals, 38 Events, 16 Locations, and 16 Organizations. When writing this paper, the domain expert was still resolving conflicts regarding some of the discovered relations, so we prefer not to report metrics on relations at this stage, leaving them for the camera-ready version.  

\subsection{Results}

To analyze how we fare on the task of recognizing entities in the graph, we count the number of Entities of each type that were recognized adequately by our automatic approach, as well as the number of entities we failed to identify and the number of entities that were extracted by the automatic approach but where not included in the gold standard graph by the domain expert. Because of the sheer number of nodes, we again resorted to LLM for this evaluation and asked LLMs to decide which nodes matched and which did not (see the Appendix). Results are presented in table \ref{Tab:Results}. 

\begin{table}[t]
\centering
\begin{tabular}{|p{2.2cm}|p{5cm}|p{0.6cm}|}
\hline
\textbf{Type of Entity} &  \multicolumn{2}{c|}{} \\ 
\hline
\multirow{3}{*}{Individual} & Present in both graphs & 47 \\ 
                         & Extra nodes not in gold standard & 0 \\ 
                         & Missing nodes from gold standard  & 2 \\ 
\hline
\multirow{3}{*}{Organization} &Present in both graphs& 14 \\ 
                         & Extra nodes not in gold standard & 21 \\ 
                         & Missing nodes from gold standard  & 2 \\ 
\hline
\multirow{3}{*}{Location} & Present in both graphs& 14 \\ 
                         & Extra nodes not in gold standard & 6 \\ 
                         & Missing nodes from gold standard  & 2 \\ 
\hline
\multirow{3}{*}{Event} & Present in both graphs& 18 \\ 
                         & Extra nodes not in gold standard & 3 \\ 
                         & Missing nodes from gold standard  & 20 \\ 
\hline
\end{tabular}
\label{Tab:Results}
\caption{comparison of the entities in the automatically generated graph and the gold standard graph.}
\end{table}

Interestingly, the precision of our algorithm varies tremendously depending on the entity's nature. The algorithm extracts individuals with high precision, and the only two missed nodes from the gold standard correspond to spouses of people who were reported to be detained and mentioned only once in all documents. 

For the case of organizations, the comparison reveals a slight problem with our algorithm, as the automatic graph creates a few extra organizations. This happens when the same organization is mentioned in the document several times with different roles (i.e., a specific division of the police is mentioned to detain someone, and then this particular division is said to be led by a specific officer). In this case, our automatic construction would generate different entities for the various natures of the organizations (so one node corresponds to the police described as an organization that detains people, and another to the police described as an organization led by this officer). 

Locations can be explained by a completely different phenomenon: our automatic graphs create a more detailed location graph than the one made by experts. For example, if an event is said to happen on the street $S$ of city $C$, experts only mention the city $C$, but our graph mentions both cities and streets. This suggests that our automatic graph could further refine the gold standard. 

Finally, we have a similar problem with locations in the case of events. Still, in the opposite direction, experts provide much more detailed information about specific events, whereas our automatic extraction tends to merge two or three events into a single one. For example, experts would detail all the events leading to the detention of victims, including the order in which individuals were detained, how they were transported, and in which vehicle; the automatic graph would instead merge all these events into a single detention event concerning all individuals, or sometimes two events, one for the detention and another for the transportation of the victims. 

\section{Conclusions and lessons learned}

The work presented here was a fully automatic analysis of a historical archive. We use large language models to recognize entities, perform entity resolution, establish relations, and resolve conflicts between identified relations. Importantly, we ensured that all the steps in our process preserve the \emph{linkage} of identified entities and relations. The final knowledge graph preserves pointers to the original documents from which a given entity or relation was extracted (note that a single entity or relation may point to several documents, as entities and relations we identify are subject to a resolution step). 

One crucial issue in this work is to measure the quality of the final knowledge graph. As several steps are involved, a slight change in the prompt used for any one step can create a massive difference in the resulting knowledge graph. Our approach was to measure the final product against a gold-standard graph. To test our algorithms, we first asked experts to manually construct a graph of 121 entities out of a small sample of documents in \url{memoriaviva.org} and test it against the result of our algorithm when we input the same subset of texts.

The resulting algorithms show promising results. Measured against the gold standard, the extraction of Individuals was extremely precise. On the other hand, the extraction of Organization, Events, and Location was less successful, but on closer inspection, this can be partially explained by differences in the granularity of the location and events, and one can verify that the automatic approach captures the big picture of events just as well as the gold standard. It remains to be seen whether this difference in granularity can be fixed by using better prompts or if we need new components in our architecture. 

In terms of future work, we plan to evaluate the entities detected by chatGPT systematically. To do so, we want to create an annotated corpus with the entities of interest \cite{fort2016collaborative}. Even if the annotation is time-consuming, it can evaluate generative AI and other possible approaches to detect critical information, such as Named Entity Recognition algorithms \cite{rojas-etal-2022-simple}.


Building a knowledge graph from collections of historical archives related to the Chilean dictatorship is a way to add robust data science analyses to this critical event in Chilean history that still has so many open questions. Unlike manual analysis of fragmented archives, a unified approach can provide a comprehensive view, filling information gaps, cross-validating data, and enhancing contextual understanding. A  holistic method can uncover deeper insights into the complexities of the dictatorship, enabling researchers to answer more nuanced questions and revealing connections and patterns that fragmented, manual analyses might miss. Of course, the quality of our analysis depends on the data quality. Thus, we are aware that further research should use primary sources and not rely on the compilation performed by \url{memoriaviva.org}.

\bibliographystyle{apalike}

\appendix
\newpage

\section{Appendix}
\subsection{Prompt for comparing entities of two graphs}

\begin{tcolorbox}[mypromptstyle, title= Prompt to extract Individuals]
\textbf{Prompt:} {\tt You will be given two lists of JSON entities extracted from the same source of information but using different methods. 
Your goal is to identify the elements of both lists that might represent totally or partially the same information, by comparing the data of both.
You will return a JSON object with a mapping from entities of the first to the second list with the following structure:}
\begin{lstlisting}[basicstyle=\ttfamily\small, breaklines=true]
[
    "map":
    [
        {
            "list_1_id": number, // id of an entity of the first list
            "list_2_id": number // id of an entity of the second list
        }
    ]
]
\end{lstlisting}
{\tt 
Consider that a node from one list might group various nodes from the other list.

Consider the following input and output example.}
\begin{lstlisting}[basicstyle=\ttfamily\small, breaklines=true]

Details ommited due to space constraints. 


\end{lstlisting}

\end{tcolorbox}
\end{document}